\documentstyle[12pt]{article}
\input epsf

\begin{document}
\baselineskip=15pt
\newcommand{\x}{{\bf x}}
\newcommand{\y}{{\bf y}}
\newcommand{\z}{{\bf z}}
\newcommand{\bp}{{\bf p}}
\newcommand{\A}{{\bf A}}
\newcommand{\B}{{\bf B}}
\newcommand{\p}{\varphi}
\newcommand{\del}{\nabla}
\newcommand{\be}{\begin{equation}}
\newcommand{\ee}{\end{equation}}
\newcommand{\bq}{\begin{eqnarray}}
\newcommand{\eq}{\end{eqnarray}}
\newcommand{\ba}{\begin{eqnarray}}
\newcommand{\ea}{\end{eqnarray}}
\def\r{\nonumber\cr}
\def\hf{\textstyle{1\over2}}
\def\qr{\textstyle{1\over4}}
\def\Sc{Schr\"odinger\,}
\def\sc{Schr\"odinger\,}
\def\'{^\prime}
\def\>{\rangle}
\def\<{\langle}
\def\-{\rightarrow}
\def\dbd{\partial\over\partial}
\def\tr{{\rm tr}}

\begin{titlepage}
\vskip1in
\begin{center}
{\large Test of gauged ${\mathcal N}=8$ SUGRA / ${\mathcal N}=1$ SYM
duality at sub-leading order.}
\end{center}
\vskip1in
\begin{center}
{\large
Paul Mansfield$^a$ and David Nolland$^b$}

\vskip.1in

{\it $^a$Department of Mathematical Sciences\\ University of Durham\\ South Road \\Durham, DH1 3LE, England}\\ {\rm e-mail: p.r.w.mansfield@durham.ac.uk}

\medskip

{\it $^b$Department of Computing and Mathematics\\ University of Northumbria\\ Ellison Building, Ellison Place\\ Newcastle Upon Tyne, NE1 8ST, England}\\ {\rm e-mail: david.nolland@unn.ac.uk}

\end{center}
\vskip1in
\begin{abstract}

\noindent
An infra-red fixed point of ${\mathcal N}=1$ super-Yang-Mills theory is
believed to be dual to a solution of five-dimensional gauged ${\mathcal N}=8$
supergravity. We test this conjecture at next to leading order in the
large $N$ expansion by computing bulk one-loop corrections to the anomaly coefficient $a-c$. The one-loop corrections are non-zero for all values of the bulk mass, and not just special ones as claimed in previous work.

\end{abstract}

\end{titlepage}

There have been many successful tests of Maldacena's conjecture \cite{Maldacena} that IIB string theory compactified on
$AdS_5\times S^5$ is dual to ${\mathcal N}=4$ super-Yang-Mills theory
with gauge group $SU(N)$ on the boundary of $AdS_5$. This has led to extensive studies of other conjectured holographic dualities.

A particularly interesting possibility is that an infra-red
fixed point of massive ${\cal N}=1$ super-Yang-Mills theory can be described
by a solution of five-dimensional gauged ${\mathcal N}=8$
supergravity, \cite{warner}, \cite{8}, \cite{9}. The purpose of the present letter is to
test this correspondence at next to leading order in the large-$N$ expansion.

The $AdS_5\times S^5$ compactification \cite{Kim} pertinent to Maldacena's original
conjecture is constructed by assuming that in
the field theory limit only the metric and five-form field strength have
non-zero vacuum expectation values. Kaluza-Klein decomposition
of the fields on $S^5$ gives an infinite number of supermultiplets of
$U(2,2/4)$ propagating in $AdS_5$. Assuming that  this
theory can be consistently truncated to its `massless' multiplet leads to
a description in terms of  five-dimensional gauged ${\mathcal
N}=8$ supergravity \cite{5},\cite{7}. Allowing certain scalars also to be
non-zero introduces mass terms into the boundary theory and breaks ${\mathcal
N}=4$ down to ${\mathcal N}=1$. This deforms the $AdS_5\times S^5$ structure
of the bulk theory, but when the scalars
are at the critical point of their potential the bulk spacetime
is again $AdS_5$ and so the boundary theory must be conformally
invariant. It is thought to be the infra-red fixed point
of the renormalisation group flow driven by the mass terms. Evidence for this
comes from a comparison of symmetries in the bulk and boundary theories,
and a comparison of the results of tree-level computations in the bulk theory,
(valid to leading order in large $N$) with exactly computable quantities in the
boundary theory. For example the bulk tree-level mass spectrum has been
compared with scaling dimensions in the boundary theory \cite{warner}, and the
trace anomaly coefficients of the boundary
theory (conventionally called $a$ and $c$) have been correctly reproduced to leading order in $N$ using the
saddle-point method of \cite{Henningson} in the bulk theory
\cite{warner},\cite{9},\cite{N}.
In this letter we show that the latter test holds also at the
next order in  the large-$N$ expansion. We consider only the combination $a-c$, as was done for the Maldacena conjecture itself in \cite{us2}.

When a four dimensional gauge theory is coupled to a non-dynamical,
external  metric, $g_{ij}$, the Weyl anomaly, ${\cal A}$, is the response of
the logarithm of the partition function, $F[g_{ij}]$,
to a scale transformation of that metric:
\be\delta F=\int d^4x\,{\sqrt g} \,
\delta\rho \,\langle T^i_i\rangle=
\int d^4x\,{\sqrt g} \,
\delta\rho\, {\cal A},\qquad
\delta g_{ij}=\delta\rho \,g_{ij},\ee
with $T_{ij}$ the stress-tensor.
On general grounds ${\cal A}$ must be a
linear combination of the Euler density, $E$, and the square of the Weyl
tensor, $I$, so ${\cal A}=a\,E+c\,I$. The coefficients $a,\, c$ are known
exactly both for the ${\mathcal N}=4$ gauge theory that is the subject of
Maldacena's conjecture and for the infra-red fixed point of the ${\mathcal N}=1$
theory. In both cases $a=c$. The numerical values of $a$ and $c$ are reproduced
to leading order in large $N$ by tree-level calculations in the appropriate
bulk supergravity theories using just the Einstein-Hilbert part of the
action. In \cite{us2} it was shown that for the Maldacena conjecture $a=c$
continued to hold at next to leading order in $N$ when bulk one-loop effects
contribute for each of the fields in the supergravity theory. The results of
\cite{us2} can be summarised in the statement  that to this order $a-c$ is
proportional to $\Phi\equiv\sum  (\Delta-2)\,\alpha$, where $\Delta$ depends
on the mass of the bulk field and  coincides with the scaling dimension of the
 boundary Green's function and  $\alpha$ is a numerical coefficient occuring
in the short-distance expansion  of an appropriate heat-kernel. The sum runs
over all the fields in the bulk  theory. The values  of $\alpha$ are: $1$ for
a real scalar, $\phi$; $7/2$ for a  complex spinor, $\chi$; $-11$ for a vector,
$A_\mu$; $33$ for a two-form,  $B_{\mu\nu}$; $-219/2$ for a complex
Rarita-Schwinger field, $\psi_\mu$;  and
$189$ for a symmetric second rank tensor, $h_{\mu\nu}$.
Faddeev-Popov ghosts must be
included when there is a gauge symmetry. This occurs for vector fields when
$\Delta-2=1$, when the net effect of the ghosts is an additional contribution
to $\Phi$ of $-2$. Similarly, the Rarita-Schwinger field has a gauge symmetry
when $\Delta-2=3/2$ and the ghosts contribute $-35/4$, and the graviton has a
gauge symmetry for $\Delta-2=2$ in which case the ghosts contribute $33$.
When $\Phi$ was computed for the Maldacena conjecture \cite{us2} it was found
that the sum over each of the infinite number of $U(2,2/4)$ supermultiplets
vanished so that $a=c$ for the full theory agreeing with the boundary theory
result.

The supergravity theory conjectured to be dual to the infra-red fixed
point of the ${\mathcal N}=1$ gauge theory is a truncation of the
ten-dimensional theory and its fields are organised into a finite number of
supermultiplets of $SU(2,2|1)$. $\Phi$
is readily computed from
the mass spectrum. The ingredients of the calculation
are given in tables 1 and 2 corresponding to tables 6.1 and 6.2 of
\cite{warner}. Each field is in a representation of
$SU(2)_I$. The dimensions of the representations are given in the tables
and contribute to $\Phi$. When the representation is complex the contribution
to $\Phi$ is doubled for the bosonic fields. The fermionic fields are
assumed to be complex already so a factor of $1/2$ is included for real
representations. We should note that
the sign of the contribution of the first scalar in Table 1
is taken in accordance with the comments of \cite{warner} so as to fit
the standard relation for the scaling dimensions of chiral primaries.
Summing $\Phi$ over all the supermultiplets in the tables
gives
$$\Phi=0-135/4+225/2+45/4+225+0-675/2+45/2=0\,,$$
so that $a=c$ to next to leading order in $N$ in the bulk theory,
in agreement with the exact result in the boundary theory
to which it is conjectured to be dual. This result also provides a check on
the spectrum of \cite{warner}.

Finally we outline the derivation of $\Phi$. It is easy to solve
Einstein's equations with cosmological constant $\Lambda=-6/l^2$
in the bulk when the boundary metric is Ricci flat to obtain \be
ds^2={1\over t^2}\left(
l^2\,dt^2+\tau^{'2}\sum_{i,j}g_{ij}\,dx^i\,dx^j\right),\qquad t\ge
\tau' \label{newmet}  \ee where $\tau'$ is a regulator that
ultimately should be taken to zero. For a Ricci-flat boundary
$E=-I$ so that ${\mathcal A}=(a-c)\,E$, and working with this
metric will only reveal the combination $a-c$. The central object
of interest in the AdS/CFT correspondence is the `partition
function' given as a functional integral for the bulk theory in
which the fields have prescribed values, $\varphi$, on the
boundary at $t=\tau'$ \cite{Witten, Klebanov}. The regulator is
necessary even in tree-level calculations but at one-loop we also
need a large $t$ cut-off $\tau$;  this introduces another
boundary, and the functional integral should be performed with the
fields taking prescribed values, $\tilde\varphi$,   there as well.
Consequently the partition function is the limit as   the cut-offs
are removed of a functional
$\Psi_{\tau,\tau'}[\tilde\varphi,\varphi]$. The exponential of
$F[g_{ij}]$ is the field independent part of this partition
function. With the regulators in place the free energy becomes a
function of $\tau,\tau'$, and the Weyl anomaly can be found by
exploiting the invariance of the five-dimensional metric
(\ref{newmet}) under $t\rightarrow (1+ \delta\rho/2)t$,
$g_{ij}\rightarrow (1+ \delta\rho)g_{ij}$, with $\delta\rho$
constant. So, for a constant Weyl scaling \be \delta F=\int
d^4x\,{\sqrt g} \, \delta\rho \,{\cal A}=-{\delta\rho\over
2}\left(\tau{\partial F\over\partial\tau}+ \tau'{\partial
F\over\partial\tau'}\right) \label{der} \ee

\begin{table}[tbp]
\begin{center}
\caption{$\Phi$ for the five short $SU(2,2|1)$ representations}
\label{coeffss}  \vskip .3cm
\begin{tabular}{|c|c|cccccc|c|}
\hline
Representation & $\Delta-2$ & $\phi$ & $\chi$ & $A_\mu$ &
$B_{\mu\nu}$ & $\psi_\mu$ & $h_{\mu\nu}$ & $\Phi$ \\  \hline \hline
${\cal D}({3/2},0,0;1)$ & $-{1/2}$ & ${\bf 3}$ &&&&&&$-3$\\
&0&&{\bf 3}&&&&& 0\\
complex & ${1/2}$ & ${\bf 3}$ &&&&&&$3$\\
\hline
Total&&&&&&&&$\bf 0$\\
\hline
\hline
 ${\cal D}({2},0,0;0)$ & $0$ & ${\bf 3}$ &&&&&&$0$\\
&${1/ 2}$&&${\bf 3 \oplus 3}$&&&&& ${21/ 4}$\\
real & ${1}$ &&&$\bf 3$&&&&$-39$\\
\hline
Total&&&&&&&&$\bf -{135/4}$\\\hline \hline
${\cal D}({9/4},1/2,0;3/2)$ & ${1/4}$ & &${\bf 2}$ &&&&&$7/4$\\
&3/4&{\bf 2}&&&$\bf 2$&&& 102\\
complex & ${5/4}$ & &${\bf 2}$ &&&&&$35/4$\\
\hline
Total&&&&&&&&$\bf 225/2$\\
\hline
\hline
${\cal D}({3},0,0;2)$ & ${1}$ & ${\bf 1}$ &&&&&&$2$\\
&3/2&&{\bf 1}&&&&& 21/4\\
complex & ${2}$ & ${\bf 1}$ &&&&&&$4$\\
\hline
Total&&&&&&&&$\bf 45/4$\\
\hline
\hline
${\cal D}({3},1/2,1/2;0)$ & $1$ &&&$\bf 1$&&&&$-13$\\
&${3/ 2}$&&&&&${\bf 1 \oplus 1}$&& $-173$\\
real & ${2}$ &&&&&&$\bf 1$&$411$\\
\hline
Total&&&&&&&&$\bf 225$\\\hline \hline

  \end{tabular}
 \end{center}
\end{table}

At one-loop we only need the quadratic fluctuations in the action,
so the fields are essentially free. In \cite{us} we computed the
Weyl anomaly for free scalar and spin-half particles for the
metric (\ref{newmet}), not by performing a functional integration
but by interpreting $\Psi_{\tau,\tau'}[\tilde\varphi,\varphi]$
(after Wick rotation of $g_{ij}$) as the \Sc functional, i.e. the
matrix element of the time evolution operator between eigenstates
of the field, $ \langle \,\tilde\varphi\,| \, T\,
\exp(-\int_{\tau'}^{\tau} d t\, H ( t) )\,|\,\varphi\,\rangle $
$=\Psi_{\tau,\tau'}[\tilde\varphi,\varphi] \,.$

\begin{table}[tbp]
\begin{center}
\caption{$\Phi$ for the remaining $SU(2,2|1)$ representations}  \label{coeffs}
\vskip .3cm
\begin{tabular}{|c|c|ccccc|c|}
\hline
Representation & $\Delta-2$ & $\phi$ & $\chi$ & $A_\mu$ &
$B_{\mu\nu}$ & $\psi_\mu$ & $\Phi$ \\  \hline \hline
 ${\cal D}({\sqrt 7+1},0,0;0)$ & $\sqrt 7-1$ & ${\bf 1}$ &&&&& $\sqrt 7-1$\\
&${\sqrt 7-1/ 2}$&&${\bf 1 \oplus 1}$&&&& ${7(\sqrt 7-1/2)/2}$\\
real & ${\sqrt 7}$ & ${\bf 1 \oplus 1}$    &&$\bf 1$&&&$-9\sqrt 7$\\
&${\sqrt 7+1/ 2}$&&${\bf 1 \oplus 1}$&&&& ${7(\sqrt 7+1/2)/2}$\\
& $\sqrt 7+1$ & ${\bf 1}$ &&&&&$ \sqrt 7+1$\\\hline
Total&&&&&&&$\bf 0$\\\hline \hline
${\cal D}({11/4},1/2,0;1/2)$ & ${3/4}$ & &${\bf 2}$ &&&&$21/4$\\
&5/4&{\bf 2}&&$\bf 2$& $\bf 2$&& $115$\\
complex & ${7/4}$ & &${\bf 2\oplus 2}$ &&&$\bf 2$&$-1435/4$\\
& ${9/4}$ &    &&$\bf 2$&&&$-99$\\
\hline
Total&&&&&&&$\bf -675/2$\\
\hline
\hline
${\cal D}({3},0,1/2;1/2)$ & ${1}$ & &${\bf 1}$ &&&&$7/2$\\
&$3/2$&&&${\bf 1}$&$\bf 1$&& $66$\\
complex & ${2}$ & &${\bf 1}$ &&&$\bf 1$&$-212$\\
&$5/2$&&&&$\bf 1$&& $165$\\
\hline

Total&&&&&&&$\bf 45/2$\\
\hline
\hline

  \end{tabular}
 \end{center}
\end{table}

To illustrate this consider a massless scalar field propagating in
the metric (\ref{newmet}). $\Psi$ satisfies the  functional
Schr\"odinger equation \be {\partial\over
\partial\tau}\Psi_{\tau,\tau'}[\tilde\varphi,\varphi] ={\textstyle
1\over 2}\int d\x\,\left
(\tau^3\,{\delta^2\over\delta\tilde\varphi^2} +\tau^{-3}\,
\tilde\varphi \nabla  \cdot\nabla\tilde\varphi
+2\,\delta^4(0)/\tau\right)\Psi_{\tau,\tau'}[\tilde\varphi,\varphi],\label{Schr}
\ee with a similar equation for the $\tau'$ dependence, and an
appropriate initial condition as $\tau'$ approaches
$\tau$.\footnote{This initial condition is quite subtle. In flat
space we would get a delta-functional in the
$\tau\rightarrow\tau'$ limit. The solution we found in \cite{us2}
has the delta-functional property in this limit provided we use
the $\epsilon$ prescription discussed here.} $\log\Psi$ has the
form $ F+\int d^d\x\,\left({1\over
2}\tilde\varphi\,\Gamma_{\tau,\tau'}\,\tilde\varphi+\tilde\varphi\,\Xi_{\tau,\tau'}\,
\varphi+{1\over 2}\varphi\,\Upsilon_{\tau,\tau'}\,\varphi\right)
$. The kernels can be expressed in terms of simpler operators
$\Gamma_{\tau,0}\equiv \Gamma(\Omega)/\tau^3$, $\Xi_{\tau,0}\equiv
\Xi(\Omega)/\tau^3$ and $\Upsilon_{\tau,0}\equiv
\Upsilon(\Omega)/\tau^3$, where $\Omega\equiv-\tau^2\nabla^2$ by
using the self-reproducing property $\int
\Psi_{\tau,\tau'}[\tilde\varphi,\phi] \,{\mathcal
D}\phi\,\Psi_{\tau',\tau''}[\phi,\varphi]=
\Psi_{\tau,\tau''}[\tilde\varphi,\varphi]$:
$$\Gamma_{\tau,\tau'}={1\over\tau^4}\left(\Gamma(\Omega)+\left(\left(
{\tau\over\tau'}\right)^4\Upsilon(\Omega')
-\Upsilon(\Omega)\right)^{-1}\Xi^2(\Omega)\right)\,,$$
$$\Upsilon_{\tau,\tau'}={1\over(\tau'+\epsilon)^{4}}\left(-\Gamma(\Omega')+\left( \Upsilon(\Omega') -\left({\tau'+\epsilon\over
\tau}\right)^4\Upsilon(\Omega)\right)^{-1}\Xi^2(\Omega')\right)\,,$$
where
$\Omega'\equiv-\tau^{'2}\nabla^2$.
The $\epsilon$ prescription is needed to ensure that this last expression
reduces to $\Upsilon(\Omega)/\tau^3$ as $\tau'\rightarrow 0$.  The \Sc equation
relates $\tau'\partial F/\partial\tau'$ to the functional trace of
$(\tau'+\epsilon)^4\Upsilon_{\tau,\tau'}$. When this is regulated with a
cut-off on $\Omega'$ then it simplifies as $\tau'\rightarrow 0$ and
$\tau\rightarrow \infty$
$$-\Gamma(\Omega')+\left(
\Upsilon(\Omega') -\left({\tau'+\epsilon\over
\tau}\right)^4\Upsilon((\tau^2/\tau^{'2})\Omega')\right)^{-1}\Xi^2(\Omega')
\rightarrow-\Gamma(\Omega')$$
because for large argument $\Upsilon((\tau^2/\tau^{'2})\Omega)\sim
((\tau^2/\tau^{'2})\Omega)^2$. Similarly
$\tau\partial F/\partial\tau$ is related to the functional trace of
$\tau^4\Gamma_{\tau,\tau'}$ which must be regulated with a cut-off on
$\Omega$ and simplifies as $\tau'\rightarrow 0$ and
$\tau\rightarrow \infty$
$$\Gamma(\Omega)+\left(\left(
{\tau\over\tau'}\right)^4\Upsilon((\tau^{'2}/\tau^2)\Omega')
-\Upsilon(\Omega)\right)^{-1}\Xi^2(\Omega)\rightarrow\Gamma(\Omega)$$
$\Gamma$ is obtained as a power series in $\Omega$ from the
\Sc equation, and the regulated traces are calculated using the heat
kernel for $\Omega$. The short distance expansion of the heat kernel finally
gives the trace as being proportional to $-\Gamma(0)=2(\Delta-2)$. This is readily extended  to massive scalars, and, with some work, to all the other
fields in the theory.

The simplification of the traces of
$\Gamma_{\tau,\tau'}$ and $\Upsilon_{\tau,\tau'}$ to that of $\Gamma$
occurs for all values of the mass, not just special values as claimed in
\cite{us2,us}. This is a result of the $\epsilon$ prescription introduced above. The prescription may be understood by writing the Schr\"odinger functional in terms of field variables which give the expected flat-space behaviour in the $\tau\rightarrow\tau^\prime$ limit. A similar prescription holds for fermions.

\vfill\eject

\end{document}